\pdfoutput=1
\documentclass[12pt,a4]{article}
\usepackage{relsize}
\usepackage{array}
\usepackage{a4wide}
\usepackage[linktoc=all,hidelinks]{hyperref}
\usepackage{latexsym}
\usepackage{cite}
\usepackage{mathrsfs}
\usepackage{amssymb}
\usepackage{amsmath}
\usepackage{graphicx}
\usepackage{enumerate}
\usepackage[usenames,dvipsnames]{xcolor}

\newcommand{\umu}{{\underline{\vphantom{\mathstrut}\mu}}}
\newcommand{\ula}{{\underline{\vphantom{\mathstrut}\lambda}}}
\newcommand{\urho}{{\underline{\vphantom{\mathstrut}\rho}}}

\newcommand{\be}{\begin{equation}\label}
\newcommand{\ee}{\end{equation}}
\newcommand{\bea}{\begin{eqnarray}\label}
\newcommand{\eea}{\end{eqnarray}}

\newcommand{\zb}{\bar{z}}

\makeatletter
\newcommand*{\textoverline}[1]{$\overline{\hbox{#1}}\m@th$}
\newcommand*\bigcdot{\mathpalette\bigcdot@{.65}}
\newcommand*\bigcdot@[2]{\mathbin{\vcenter{\hbox{\scalebox{#2}{$\m@th#1\bullet$}}}}}
\makeatother

\date{}
\begin{document}

\title{Recursion Relations for Anomalous Dimensions in the 6d $(2,0)$ Theory}
\author{Theresa Abl, Paul Heslop and  Arthur E. Lipstein \vspace{7pt}\\ \normalsize \textit{
Department of Mathematical Sciences}\\\normalsize\textit{Durham University, Durham, DH1 3LE, United Kingdom}}
\maketitle
\begin{abstract}
\noindent We derive recursion relations for the anomalous dimensions of double-trace operators occurring in the conformal block expansion of four-point stress tensor correlators in the 6d $(2,0)$ theory, which encode higher-derivative corrections to supergravity in $AdS_7 \times S^4$ arising from M-theory. As a warm-up, we derive analogous recursion relations for four-point functions of scalar operators in a toy non-supersymmetric 6d conformal field theory. 

\end{abstract}
\thispagestyle{empty}
\pagebreak
\tableofcontents

\section{Introduction}
Understanding the stable 5-dimensional objects of M-theory known as M5-branes is one of the most important open questions in string theory. A stack of coincident M5-branes in flat background can be described by a 6d superconformal theory with $(2,0)$ supersymmetry which is dual to M-theory in $AdS_7 \times S^4$, although very little is known about the model since it is intrinsically strongly coupled. A lot of progress has been made by dimensionally reducing the theory or computing quantities protected by supersymmetry, but ultimately one wants to compute unprotected quantities in six dimensions. In this regard, a very promising strategy is the conformal bootstrap, which aims to use the operator product expansion (OPE) and crossing symmetry of four-point correlators to fix the OPE coefficients and scaling dimensions of the theory (collectively known as the OPE data) \cite{Ferrara:1973yt,Ferrara:1973vz,Polyakov:1974gs,Rattazzi:2008pe}. This approach was first applied to the 6d $(2,0)$ theory in \cite{Beem:2015aoa}.

In this paper, we analyse four-point correlation functions of stress tensor multiplets in the 6d $(2,0)$ theory in the limit of large central charge, $c$. In~\cite{Heslop:2017sco} the form of the first corrections  to the correlator (due to higher-derivative terms in the M-theory action) were found. The method there was to conjecture the form of suitable crossing symmetric functions and then check that the conformal block expansions of the resulting functions had the expected spin truncation. The main aim of this paper is to derive these spin truncated solutions directly from the crossing equations, confirming the results of~\cite{Heslop:2017sco} as well as giving an alternative and more direct method for obtaining higher-derivative results.

Our main strategy, adapted from the seminal work of \cite{Heemskerk:2009pn}, is to expand the crossing equations in the inverse central charge and then take a certain limit of the conformal cross ratios to isolate the terms in the conformal block expansion corresponding to anomalous dimensions of double-trace operators. We then truncate the conformal block expansion in spin and use the orthogonality of the hypergeometric functions in the superconformal blocks to derive a recursion relation for the anomalous dimensions. For truncated spin $L$, we find that the solution to the recursion relation depends on $(L+2)(L+4)/8$ free parameters, in agreement with holographic arguments of \cite{Heemskerk:2009pn} and with the explicit four-point functions found in~\cite{Heslop:2017sco}. In particular, they can be thought of as the coefficients of higher-derivative corrections to supergravity in $AdS_7 \times S^4$ arising from M-theory \cite{Heslop:2017sco} (see \cite{Alday:2014tsa,Goncalves:2014ffa} for similar results in $\cal N$=4 super Yang-Mills theory). 

A strategy for fixing the coefficients using a chiral algebra conjecture \cite{Beem:2014kka} was recently proposed in \cite{Chester:2018dga}. Moreover, the M-theory effective action can also be deduced from correlators of the ABJM theory \cite{Aharony:2008ug}, which is dual to M-theory in $AdS_4 \times S^7$ \cite{Chester:2018lbz,Chester:2018aca}. As a warm-up for our analysis in the $(2,0)$ theory, we first derive recursion relations for anomalous dimensions in an abstract non-supersymmetric 6d conformal field theory (CFT), which we match against the conformal block expansion of Witten diagrams for a massive scalar field in $AdS_7$. The recursion relations we obtain for this toy model and the $(2,0)$ theory can be efficiently solved using a computer, and we include the Mathematica file {\tt{6drecursion.nb}} for doing so. 

The structure of this paper is as follows. In section~\ref{toy} we derive recursion relations for anomalous dimensions in a toy 6d model and match the solutions against the conformal block expansion of Witten diagrams in $AdS_7$. In section~\ref{susy}, we then adapt this analysis to the 6d $(2,0)$ theory, and match the solutions of the supersymmetric recursion relations with the results obtained in \cite{Heslop:2017sco}. In section~\ref{conclusion} we present our conclusions and future directions. There are also several appendices. In appendix~\ref{blocks}, we provide formulas for the conformal blocks in terms of hypergeometric functions and in appendix~\ref{orthogonality} we derive inner products for these functions. In appendix~\ref{algorithm} we describe a general algorithm for solving the recursion relations for anomalous dimensions, and in appendix~\ref{SolsSpin2} we describe the solutions for spin truncation $L=2$.

\section{Toy Model} \label{toy}

In \cite{Heemskerk:2009pn} the authors considered four-point correlators of scalar operators in an abstract non-supersymmetric CFT in two and four dimensions, and showed that the solutions to the crossing equations whose conformal block expansion is truncated in spin are in one-to-one correspondence with local quartic interactions of a massive scalar field in $AdS$ (modulo integration by parts and equations of motion). In this section, we will carry out a similar analysis for a toy model in six dimensions as a warm up for our analysis of the 6d $(2,0)$ theory in the next section. In particular, we will analyse four-point correlators of a scalar operator $\mathcal{O}$ with classical dimension $\Delta_0$. A four-point correlator then has the form \cite{Dolan:2003hv}
\begin{equation}
\left\langle \mathcal{O}_{1}\mathcal{O}_{2}\mathcal{O}_{3}\mathcal{O}_{4}\right\rangle =\frac{F(u,v)}{\left(x_{12}^{2}\right)^{\Delta_{0}}\left(x_{34}^{2}\right)^{\Delta_{0}}},
\end{equation}
where $\vec{x}_{i}$ is the position of the $i$'th operator, $x_{ij}^{2}=\left(\vec{x}_{i}-\vec{x}_{j}\right)^{2}$, and $F$ is a function of the conformal cross ratios  
\begin{equation}\label{zzb}
u=\frac{x_{12}^{2}x_{34}^{2}}{x_{13}^{2}x_{24}^{2}}=z\bar{z},\qquad v=\frac{x_{14}^{2}x_{23}^{2}}{x_{13}^{2}x_{24}^{2}}=(1-z)(1-\bar{z}),
\end{equation}
where we will use the variables $(u,v)$ interchangeably with $(z,\bar{z})$. Note that exchanging $\vec{x}_2$ with $\vec{x}_4$ corresponds to exchanging $u$ and $v$, or $(z,\bar{z})$ with $(1-z,1-\bar{z})$. Invariance of the correlator under this exchange (known as crossing symmetry) then implies the following constraint on $F$:
\begin{equation}\label{crossing1}
v^{\Delta_{0}}F(u,v)=u^{\Delta_{0}}F(v,u).
\end{equation}
In this model, the primary double-trace operators are schematically
\begin{equation}
\mathcal{O}_{n,l}=\mathcal{O}\partial_{\mu_{1}}...\partial_{\mu_{l}}\partial^{2n}\mathcal{O},
\label{primaryb}
\end{equation}
which have scaling dimension $\Delta=2n+l+2\Delta_0+\mathcal{O}(1/c)$, spin $l$ and naive twist $2n+2\Delta_0$. 
The conformal block expansion of $F(u,v)$ is then given by the following sum over primary operators: 
\begin{equation}\label{blockexp}
F(u,v)=\sum_{n,l\geq 0}A_{n,l}\,G^{\text{B}}_{\Delta,l}(z,\bar{z}),
\end{equation}
where $A_{n,l}$ are OPE coefficients and $G^{\text{B}}_{\Delta,l}$ are the bosonic conformal blocks given in terms of hypergeometric functions in appendix~\ref{blocks}, which implicitly depend on $n$ through the scaling dimensions of the conformal primary operator $\Delta$. Note that $A_{n,l}=0$ when $l$ is odd since operators with an odd number of derivatives in the OPE of two identical operators correspond to descendants.

The free disconnected part of the four-point correlator is given by
\begin{equation}
F(u,v)_{\text{free-disc}}=1+\frac{1}{u^{\Delta_{0}}}+\frac{1}{v^{\Delta_{0}}}
\end{equation}
and its conformal block expansion gives the leading contribution to the OPE coefficients 
\begin{align}\label{A0bose}
A_{n,l}^{(0)}=&\,\frac{2\, \left(l+2\right) \left(2\Delta_0+l+2n-2\right)\left(2 \Delta_0 +l+2 n-3\right) \left(\left(\Delta_0 +n-3\right)!\right)^2 }{\left(\left(\Delta_0 -3\right)!\right)^2 \left(\left(\Delta_0 -1\right)!\right)^2 n!\left(l+n+2\right)! \left(2 \Delta_0 +2 n-6\right)! \left(2 \Delta_0 +2l+2n-2\right)!}\nonumber \\
&\times \left(\left(\Delta_0 +l+n-1\right)!\right)^2 \left(2 \Delta_0 +n-6\right)! \left(2 \Delta_0 +l+n-4\right)! 
\end{align}
In the next subsection, we will derive recursion relations for the anomalous dimensions $\gamma_{n,l}$. After solving the recursion relations, we can then deduce the $1/c$ correction to the OPE coefficients $A_{n,l}^{(1)}$ using the following formula:
\begin{equation}
A_{n,l}^{(1)}=\frac{1}{2}\partial_{n}\left(A_{n,l}^{(0)}\gamma_{n,l}\right).
\end{equation}
This formula was first found in two and four dimensions \cite{Heemskerk:2009pn,Fitzpatrick:2011dm} and was subsequently observed to hold in six dimensions \cite{Heslop:2017sco}. 

\subsection{Recursion}

In this subsection, we will derive a formula for the anomalous dimensions of double-trace operators in the toy 6d CFT described above following the method developed for 2d and 4d CFT's in \cite{Heemskerk:2009pn}. This formula will be written as a sum over the spin of the operators and will depend on two non-negative integers $p$ and $q$. Truncating the sum over spin to maximum spin $L$ and choosing $p$ and $q$ appropriately will then give rise to recursion relations for the anomalous dimensions, which can be solved for arbitrary twist and spin $l\leq L$ in terms of $(L+2)(L+4)/8$ free parameters, in agreement with counting of solutions in lower dimensions and holographic arguments, as we will describe in the next subsection. 

The first step is to expand the OPE data in $1/c$:
\begin{equation}
A_{n,l}=A_{n,l}^{(0)}+\frac{1}{c}A_{n,l}^{(1)}+...,\qquad \Delta=2n+l+2\Delta_0+\frac{1}{c}\gamma_{n,l}+...
\end{equation}
Expanding the conformal block decomposition~\eqref{blockexp} in $1/c$ and inserting this into the crossing equation~\eqref{crossing1} then gives%
\footnote{\label{f1}Note that in general there will be degeneracy in the free theory, so more than one operator with each given naive dimension and spin, so in \eqref{crossingexpansionbos} the free conformal block coefficient gives a sum over these operators of three-point coefficients squared,  $A_{n,l}^{(0)}=\sum_i \langle \Delta_0\Delta_0 i\rangle^2$. Then $\gamma_{n,l}$ is in reality the so-called ``averaged anomalous dimension''  $\gamma_{n,l}= (\sum_i \langle \Delta_0\Delta_0i\rangle^2\,\gamma_{n,l,i})/(\sum_i \langle \Delta_0\Delta_0i\rangle^2)$ where $\gamma_{n,l,i}$ are the anomalous dimensions of the individual operators. To obtain the individual anomalous dimensions requires more data, for example four-point functions of operators with different dimensions.}
\begin{equation}
v^{\Delta_{0}}\sum_{n,l\geq 0}\left[A_{n,l}^{(1)}\,G_{\Delta,l}^{\text{B}}(z,\bar{z})+\frac{1}{2}A_{n,l}^{(0)}\,\gamma_{n,l}\,\partial_{n}G_{\Delta,l}^{\text{B}}(z,\bar{z})\right]-(u\leftrightarrow v)=0.
\label{crossingexpansionbos}
\end{equation}
Note that the conformal blocks are given as a sum of products of hypergeometrics with the schematic form 
\begin{equation}\label{blockschematic}
G_{\Delta,l}^{\text{B}}(z,\bar{z})\sim\sum \frac{u^{n}}{\lambda^{3}}k_{\alpha}(z)k_{\beta}(\bar{z}),
\end{equation}
where $\lambda=z-\bar{z}$ and
\begin{equation}\label{eq:hypergeomBos}
k_\beta(z)={}_2 F_1\left(\beta/2,\beta/2,\beta,z\right)
\end{equation}
(see appendix~\ref{blocks} for the exact form of the blocks). From this we see that $\partial_{n}G^{\text{B}}_{\Delta,l}(z,\bar{z})$ gives a contribution of the form $\log(u)=\log(z \bar{z})$, and the analogous term in the cross channel will contribute $\log((1-z) (1-\bar{z}))$. As a result, we can isolate the terms containing the anomalous dimensions in both channels simultaneously by taking the $\log(z)\log(1-\zb)$ coefficient of the crossing equation as $z \rightarrow 0$ and $\bar{z} \rightarrow 1$. In order for the crossing equation to be consistent, the $\log(z)$ coming from $\partial_{n}G^{\text{B}}_{\Delta,l}(z,\bar{z})$ must thus be accompanied by a $\log(1-\bar{z})$. Such terms indeed arise from the hypergeometrics depending on $\bar{z}$ after making use of the relation
\begin{equation}
k_\beta(\zb)=\log(1-\zb)\,\tilde{k}_\beta(1-\zb)+\text{holomorphic at }\zb=1,
\label{hypertolog}
\end{equation}
where
\begin{equation}
\tilde{k}_\beta(z)=-\frac{\Gamma(\beta)}{\Gamma(\beta/2)^2}  {}_2F_1 \left(\beta/2,\beta/2,1,z\right).
\end{equation}
Similarly, the hypergeometrics depending on $1-z$ in the cross channel will give rise to $\log(z)$.

In summary, we take the $\log(z)\log(1-\zb)$ coefficient of~\eqref{crossingexpansionbos} as $z \rightarrow 0$ and $\bar{z} \rightarrow 1$  yielding the refined crossing equation:
\begin{align}
&v^{\Delta_{0}}\sum_{n,l\geq 0}A_{n,l}^{(0)}\,\gamma_{n,l}\left(\left.\partial_{n}G^{\text{B}}_{2n+l+2\Delta_{0},l}(z,\bar{z})\right)\right|_{\log z\log(1-\bar{z})}=\nonumber \\
&u^{\Delta_{0}}\sum_{n,l\geq 0}A_{n,l}^{(0)}\,\gamma_{n,l}\left(\left.\partial_{n}G^{\text{B}}_{2n+l+2\Delta_{0},l}(1-z,1-\bar{z})\right)\right|_{\log z\log(1-\bar{z})},
\label{crossing}
\end{align}into which we insert (the precise forms of) ~\eqref{blockschematic} and~\eqref{hypertolog} to obtain sums of  terms of the form $k_\alpha(z) \tilde k_\beta(1-\bar z)$ and $k_\alpha(1-\bar z) \tilde k_\beta( z)$. To extract a purely numerical recursion relation we then multiply the resulting equation by
\begin{equation}
\frac{k_{-2q}(z)}{z^{5-\Delta_0 +q}}\, \times \frac{k_{-2p}(1-\zb)}{(1-\zb)^{5-\Delta_0+p}},
\end{equation}
where $p$ and $q$ are arbitrary non-negative integers, and perform the contour integrals $\oint \frac{dz}{2\pi i} \oint \frac{d\zb}{2\pi i}$, where the contours encircle $(z,\bar{z})=(0,1)$. Using the orthogonality of the hypergeometrics obtained in \cite{Heemskerk:2009pn} (for more details see appendix~\ref{orthogonality})
\begin{equation}
\delta_{m,m'}=\oint \frac{d z}{2\pi i}\,z^{m-m'-1}\,k_{2m+4}(z)\,k_{-2m'-2}(z),
\label{boseorthogonal}
\end{equation}
and defining the integral
\begin{equation}
\mathcal{I}_{m,m'}=\oint \frac{d z}{2\pi i}\,\frac{(1-z)^{m-\Delta_0+3}}{z^{m'-\Delta_0+5}}\,\tilde{k}_{2m}(z)\,k_{-2m'}(z),
\end{equation} 
we finally arrive at the following equation:
\begin{align}\label{eq:BosRecursion}
0\,=&\sum_{l=0}^{L}\sum_{n=0}^{\infty}
A^{(0)}_{n,l}\gamma_{n,l}
\Bigl[ (l+1)\left(\delta_{q,l+n+3}\mathcal{I}_{\Delta_0 +n-3,p+\Delta_0-4}-\delta_{q,n} \mathcal{I}_{\Delta_0 +l+n,p+\Delta_0-4}\right) \nonumber\\
&+\left(l+3\right)\left(\delta_{q,n+1} \mathcal{I}_{\Delta_0 +l+n-1,p+\Delta_0-4}-\delta_{q,l+n+2}\mathcal{I}_{\Delta_0 +n-2,p+\Delta_0-4} \right)\nonumber \\
&+P_{n,l}\left(\delta_{q,l+n+3}\mathcal{I}_{\Delta_0 +n-1,p+\Delta_0-4} -\delta_{q,n+2} \mathcal{I}_{\Delta_0 +l+n,p+\Delta_0-4}\right)\nonumber \\
& +\,Q_{n,l}\left(\delta_{q,n+1} \mathcal{I}_{\Delta_0 +l+n+1,p+\Delta_0-4}-\delta_{q,l+n+4}\mathcal{I}_{\Delta_0 +n-2,p+\Delta_0-4}\right) -\left(q\leftrightarrow p\right)\Bigr],
\end{align}
where 
\begin{align}
P_{n,l}&=\frac{\left(l+3\right) \left(\Delta_0 +n-2\right)^2 \left(2 \Delta_0 +l+2 n-4\right) }{4 \left(2 \Delta_0 +2 n-5\right) \left(2 \Delta_0 +2 n-3\right) \left(2 \Delta_0 +l+2 n-2\right)},\nonumber \\
Q_{n,l}&=\frac{\left(l+1\right) \left(\Delta_0 +l+n\right)^2 \left(2 \Delta_0 +l+2 n-4\right)}{4 \left(2 \Delta_0 +l+2 n-2\right) \left(2 \Delta_0 +2 l+2 n-1\right) \left(2 \Delta_0 +2 l+2 n+1\right)}.
\end{align}
Note that we have truncated the sum over spins in~\eqref{eq:BosRecursion} to a maximum spin $L$. Recursion relations for the anomalous dimensions are then obtained by making particular choices of $p$ and $q$, and the solutions are labelled by $L$. In the next section we will explain how to solve the recursion relations for $L=0,2$ and describe the general algorithm in appendix~\ref{algorithm}.

\subsection{Solutions}\label{toysolutions}

Let us first consider the $L=0$ spin truncation in \eqref{eq:BosRecursion}. In this case, setting $q=0$ leads to the following recursion relation in terms of $p$:
\begin{equation}
\mathcal{I}_{\Delta_0,p+\Delta_0-4}A^{(0)}_{0,0}\gamma_{0,0}=\sum_{a=0}^4 C_a A^{(0)}_{p-a,0}\gamma_{p-a,0},
\end{equation}
where
\begingroup
\allowdisplaybreaks
\begin{align}
C_0=\,&\,\mathcal{I}_{p+\Delta_0,\Delta_0 -4},\nonumber \\
C_1=\,&-3\,\mathcal{I}_{p+\Delta_0-2,\Delta_0 -4}-\frac{\left(p+\Delta_0-1\right)^2 \left(p+\Delta_0-3\right)\,\mathcal{I}_{p+\Delta_0,\Delta_0 -4}}{4\left(p+\Delta_0-2\right) \left(2 p+2\Delta_0-1\right)\left(2 p+2\Delta_0-3\right)},\nonumber \\ 
C_2=\,&3\,\mathcal{I}_{p+\Delta_0-4,\Delta_0 -4}+\frac{3\left(p+\Delta_0-4\right)^3\,\mathcal{I}_{p+\Delta_0-2,\Delta_0 -4}}{4\left(p+\Delta_0-3\right)  \left(2 p+2\Delta_0-7\right)\left(2 p+2\Delta_0-9\right)},\nonumber \\
C_3=\,&-\mathcal{I}_{p+\Delta_0-6,\Delta_0-4}-\frac{3\left(p+\Delta_0-5\right)^3\,\mathcal{I}_{p+\Delta_0-4,\Delta_0 -4}}{4\left(p+\Delta_0-4\right)  \left(2 p+2\Delta_0-9\right)\left(2 p+2\Delta_0-11\right)},\nonumber \\
C_4=\,&\frac{\left(p+\Delta_0-4\right)^2 \left(p+\Delta_0-6\right)\,\mathcal{I}_{p+\Delta_0-6,\Delta_0 -4}}{4\left(p+\Delta_0-5\right)  \left(2 p+2\Delta_0-7\right)\left(2 p+2\Delta_0-9\right)}.
\end{align}\\
\endgroup
This recursion relation can be solved for all $\gamma_{n,0}$ with $n>0$ in terms of $\gamma_{0,0}$ as follows:
\begin{align}\label{bos0}
\gamma_{n,0}^{\text{spin-0}}(\Delta_0)\,=&\,\gamma_{0,0}\,\frac{\left(2 \Delta_0 -3\right) \left(2 \Delta_0 -1\right) \left(n+1\right) \left(n+2\right)  \left(\Delta_0 +n-2\right) \left(\Delta_0 +n-1\right) }{8\left(\Delta_0 -2\right)^2 \left(\Delta_0 -1\right) \left(2 \Delta_0 +2 n-5\right) \left(2 \Delta_0 +2 n-3\right)}\nonumber \\
&\times \frac{\left(2 \Delta_0 +n-5\right) \left(2 \Delta_0 +n-4\right)}{ \left(2 \Delta_0 +2 n-1\right)},
\end{align}
where we divided by $A^{(0)}_{n,0}$, see~\eqref{A0bose}. For $L=2$, first choose $(p,q)=(1,0)$ to obtain $\gamma_{1,0}$ in terms of three unfixed parameters $\left\{ \gamma_{0,0},\gamma_{0,2},\gamma_{1,2}\right\} $. For $p>1$, one can then solve the equations with $q \in \left\{ 0,1\right\} $ for $\gamma_{p,l}$ with $l \in \left\{ 0,2\right\} $ in terms of $\gamma_{p',l'}$ with $p' <p$ and $l' \in \left\{ 0,2\right\}$. In the end, we obtain a solution for all $\gamma_{n,l}$ with $l \in \left\{ 0,2\right\} $ in terms of $\left\{ \gamma_{0,0},\gamma_{0,2},\gamma_{1,2}\right\} $. The solution is a bit lengthy, so we present it in appendix~\ref{SolsSpin2}. 

An algorithm for solving the recursion relations for general spin truncation is described in appendix~\ref{algorithm} and implemented in the attached Mathematica file {\tt{6drecursion.nb}}. For a spin-$L$ truncation, we find that the solution depends on $(L+2)(L+4)/8$ unfixed parameters, in agreement with the holographic arguments of \cite{Heemskerk:2009pn}. In particular, that paper considered a massive scalar field in $AdS$ with local quartic interactions (which can be thought of as a toy model for the low energy effective action of quantum gravity in $AdS$) and showed that up to integration by parts and equations of motion, there are $L/2+1$ independent interactions which can create or annihilate a state of at most spin $L$, with the total number of derivatives ranging from $2 L$ to $3L$ in intervals of two. These can be written 
\begin{equation}
\left(\nabla^{L/2}_\umu \phi\right) \, \left(\nabla^{L/2}_\ula \phi\right) \, \left(\nabla^{k}_\urho \phi\right) \, \left(\nabla^{L+k}_{\umu \ula \urho }\phi\right) \qquad k=0,1,\dots,L/2\ ,
\label{contactterms}
\end{equation}
where the underscores denote sets of Lorentz indices. Here note that the first two scalars in isolation have $L$ free Lorentz indices as do the last two and so they can create a spin $L$ state. Hence, there is one spin-0 interaction vertex $\phi^{4}$, and two spin-2 interaction vertices equivalently written $\phi^{2}\left(\nabla_{\mu}\nabla_{\nu}\,\phi\right)^{2}$ and $\phi^{2}\left(\nabla_{\mu}\nabla_{\nu}\nabla_{\rho}\,\phi\right)^{2}$ which contain four and six derivatives, respectively. The total number of interactions up to spin $L$ is then given by $\sum_{l=0}^{L/2}(l+1)=(L+2)(L+4)/8$. 

Thus, the unfixed parameters in the solutions to the recursion relations can be identified with coefficients of the bulk interaction vertices. Indeed, we have verified that the solution in~\eqref{bos0} reproduces the anomalous dimensions in the conformal block expansion of a Witten diagram for a $\phi^4$ interaction
\begin{equation}
F^\text{spin-0}(u,v)=C^{(0)}\bar{D}_{\Delta_0\, \Delta_0\,\Delta_0\, \Delta_0}(u,v)
\end{equation}
for the following choice of free parameter:
\begin{equation}
\gamma_{0,0}=-\frac{ C^{(0)} \left(\left(\Delta_0 -1\right)!\right)^4 }{\left(2 \Delta_0 -1\right)!},
\end{equation}\\
where the coefficient $C^{(0)}$ is unfixed and the definition of $\bar{D}$ functions can be found in appendix~D of \cite{Arutyunov:2002fh}. Note that the anomalous dimensions of $F^\text{spin-0}$ are obtained by expanding this function according to~\eqref{crossingexpansionbos}. 

Moreover, the $L=2$ solution in appendix~\ref{SolsSpin2} encodes the anomalous dimensions in the conformal block expansion of Witten diagrams with four and six-derivative interactions
\begin{align}
F_4^\text{spin-2}(u,v)=& \,C^{(2)}_4 (1+u+v)\,\bar{D}_{\Delta_0 +1\,\Delta_0 +1\,\Delta_0 +1\,\Delta_0 +1}(u,v),\\
F_6^\text{spin-2}(u,v)=&\, C^{(2)}_6\left(\bar{D}_{\Delta_0 +2\,\Delta_0 +1\,\Delta_0 +2\,\Delta_0 +1}(u,v)+\bar{D}_{\Delta_0 +1\,\Delta_0 +2\,\Delta_0 +1\,\Delta_0 +2}(u,v)\right.\nonumber \\
&+u^2\,\bar{D}_{\Delta_0 +2\,\Delta_0 +2\,\Delta_0 +1\,\Delta_0 +1}(u,v)+u\,\bar{D}_{\Delta_0 +1\,\Delta_0 +1\,\Delta_0 +2\,\Delta_0 +2}(u,v)\nonumber \\
&\left.+v^2\,\bar{D}_{\Delta_0 +1\,\Delta_0 +2\,\Delta_0 +2\,\Delta_0 +1}(u,v)+v\,\bar{D}_{\Delta_0 +2\,\Delta_0 +1\,\Delta_0 +1\,\Delta_0 +2}(u,v)\right)
\end{align}
for the following choice of free parameters:
\begin{align}
\left\{ \gamma_{0,0},\gamma_{0,2},\gamma_{1,2}\right\} _{4}=&\,C_4^{(2)}\,\left\{-\frac{4\left(\Delta_0 !\right)^3 \left(\Delta_0 +1\right)!}{\left(2 \Delta_0 +2\right)!},-\frac{2\,\Delta_0 ! \left(\left(\Delta_0 +1\right)!\right)^2 \left(\Delta_0 +2\right)!}{3\left(2 \Delta_0 +1\right) \left(2 \Delta_0 +4\right)!}\right.,\nonumber \\
&\left.-\frac{\left(\Delta_0 +1\right) \left(2 \Delta_0 -1\right) \left(\Delta_0 -1\right)! \left(\left(\Delta_0 +2\right)!\right)^2 \left(\Delta_0 +3\right)!}{3\left(2 \Delta_0 +3\right) \left(2 \Delta_0 +6\right)!}\right\}, \label{bosparams1}
\end{align}
\begin{align}
\left\{ \gamma_{0,0},\gamma_{0,2},\gamma_{1,2}\right\} _{6}=&\,C_6^{(2)}\,\left\{-\frac{4\left(\Delta_0 !\right)^2 \left(\left(\Delta_0 +1\right)!\right)^2}{\left(2 \Delta_0 +2\right)!},\right.\nonumber \\
&-\frac{2 \,\left(3 \Delta_0 +2\right) \Delta_0 ! \left(\left(\Delta_0 +1\right)!\right)^2 \left(\Delta_0 +2\right)!}{3\left(2 \Delta_0 +1\right) \left(2 \Delta_0 +4\right)!},\nonumber \\
&\left.-\frac{\left(\Delta_0 +1\right) \left(6 \Delta_0 ^2+7 \Delta_0 -2\right) \left(\Delta_0 -1\right)! \left(\left(\Delta_0 +2\right)!\right)^2 \left(\Delta_0 +3\right)!}{3\left(2 \Delta_0 +3\right) \left(2 \Delta_0 +6\right)!}\right\},
\label{bosparams2} 
\end{align}
where the coefficients $C^{(2)}_{4,6}$ are unfixed. 
Note that the number of derivatives in the bulk interactions can be read off from the large-twist behaviour of the corresponding anomalous dimensions. Indeed, the anomalous dimensions of $F^\text{spin-0}$ scale like $n^3$, while those of $F_{4}^\text{spin-2}$ and $F_{6}^\text{spin-2}$ scale like $n^7$ and $n^9$, respectively.  In other words, the anomalous dimensions associated with four and six-derivative interactions scale like $n^4$ and $n^6$ compared to those of the $\phi^4$ interaction. Moreover, the ratio of the coefficients $\gamma_{0,2}$ and $\gamma_{1,2}$ in \eqref{bosparams1} can be deduced from the large-twist behaviour: for a generic choice of free parameters the anomalous dimensions scale like $n^9$ but this is reduced to $n^7$ when  $\gamma_{0,2}$ and $\gamma_{1,2}$ are related as in \eqref{bosparams1}. For more details, see appendix~\ref{SolsSpin2}.

\section{$(2,0)$ Theory} \label{susy}

In this section we will adapt the analysis of the previous section to four-point stress tensor correlators of the 6d $(2,0)$ theory. The stress tensor belongs to a half-BPS multiplet whose superconformal primary, $T_{IJ}$, is a dimension-4 scalar in the two-index symmetric traceless representation of the R-symmetry group $SO(5)$, so it is sufficient to consider four-point correlators of this operator. The abelian theory consists of a two-form gauge field with self-dual field strength, eight fermions, and five scalars $\phi^I$ \cite{Howe:1983fr,Perry:1996mk,Pasti:1997gx}, in terms of which $T_{IJ}=\phi_{(I}\phi_{J)}-\frac{1}{5}\delta_{IJ}\phi^{K}\phi_{K}$. Although it is unclear how to formulate the non-abelian theory, the $AdS$/CFT correspondence predicts that it is dual to M-theory in $AdS_7\times S^4$ and reduces to 11d supergravity in this background in the limit of large central charge \cite{Maldacena:1997re}\footnote{Using holographic methods, it has been shown that the central charge $c$ scales like $N^3$, where $N$ is the number of M5-branes \cite{Henningson:1998gx}.}. After dimensionally reducing this background on the sphere, one obtains a Kaluza-Klein tower of scalars in $AdS_7$ with masses $m_k^2=4k(k-3)$ in units of the inverse $AdS$ radius \cite{Bastianelli:1999en}, which are dual to half-BPS scalar operators in the $k$-index symmetric traceless representation of the R-symmetry group with scaling dimension $2k$. The operators we consider in this paper correspond to $k=2$ and correspond to the bottom of this tower.

Four-point correlators of stress tensor multiplets were computed in the supergravity approximation in \cite{Arutyunov:2002ff}, and a conformal block decomposition of these results was subsequently carried out in \cite{Heslop:2004du}. More recently, corrections to the supergravity approximation were deduced in \cite{Heslop:2017sco} by constructing solutions to the crossing equations whose conformal block expansion is truncated in spin. In this section, we will derive recursion relations for the (averaged) anomalous dimensions appearing in the conformal block expansions of these solutions. These recursion relations allow one to directly compute the OPE data of these solutions without having to know them explicitly, and can be straightforwardly implemented on a computer.

As shown in \cite{Dolan:2004mu,Heslop:2004du}, superconformal symmetry constrains the four-point function of stress tensor multiplets in the 6d $(2,0)$ theory in terms of a prepotential $F(z,\bar{z})$ as follows:
\begin{equation}
\lambda^{4}\left(g_{13}g_{24}\right)^{-2}\left\langle T_{1}T_{2}T_{3}T_{4}\right\rangle =\mathcal{D}\left(\mathcal{S}F\left(z,\bar{z}\right)\right)+\mathcal{S}_{1}^{2}F\left(z,z\right)+\mathcal{S}_{2}^{2}F\left(\bar{z},\bar{z}\right),
\label{prepot}
\end{equation}
where $\mathcal{D}=-\left(\partial_{z}-\partial_{\bar{z}}+\lambda\partial_{z}\partial_{\bar{z}}\right)\lambda$, the variables $z,\bar{z}$ are defined in terms of the space-time cross ratios~\eqref{zzb} and $\lambda=z-\bar z$. We have introduced auxiliary variables $Y^I$ to soak up the $SO(5)$ indices of $T_{IJ}$ via $T_i=T_{IJ}Y_i^IY_i^J$. Using these internal coordinates, we then define superpropagators $g_{ij} = Y_i \cdot Y_j / x_{ij}^4$ and internal conformal cross ratios
\begin{equation}
y\bar{y}=\frac{Y_{1}\cdot Y_{2} Y_{3} \cdot Y_{4}}{Y_{1} \cdot Y_{3}Y_{2} \cdot Y_{4}},\qquad\left(1-y\right)\left(1-\bar{y}\right)=\frac{Y_{1} \cdot Y_{4}Y_{2} \cdot Y_{3}}{Y_{1} \cdot Y_{3} Y_{2} \cdot Y_{4}},
\end{equation}
in terms of which we define $\mathcal{S}_{1}=\left(z-y\right)\left(z-\bar{y}\right)$, $\mathcal{S}_{2}=\left(\bar{z}-y\right)\left(\bar{z}-\bar{y}\right)$, and $\mathcal{S}=\mathcal{S}_{1}\mathcal{S}_{2}$.

Crossing symmetry implies that
\begin{equation}
F(u,v)=F(v,u).
\label{susycrossing}
\end{equation}
Moreover, we can write $F(u,v)$ as
\begin{equation}
F(z,\bar z)=\frac{A}{u^2}+\frac{g(z) - g(\bar z)}{u\,\lambda}+\lambda\,G(z,\bar z),
\label{decomposition}
\end{equation}
where each function in the decomposition encodes certain contributions to the OPE. Roughly speaking, $A$ encodes the unit operator, $g$ encodes protected operators, and $G$ encodes non-protected double-trace operators, which will be our main interest. In more detail, these operators have the schematic form $T\partial^{l}\Box^{n}T$ with $n \geq 0$ and scaling dimension $\Delta=2n +l +8+\mathcal{O}(1/c)$, and contribute to the conformal block expansion of $G$ as follows \footnote{Note that the conformal block expansion of $G$ also contains protected double-trace operators, which correspond to $n \in \left\{ -1,-2\right\} $ in our conventions, but we will not need to consider these operators. For more details, see \cite{Heslop:2004du}.}
\begin{equation}\label{eq:bigg}
\lambda^2 G(z,\bar z)=\sum_{n,l\geq 0} A_{n,l}\, G^{\text{S}}_{\Delta,l}(z,\bar z),
\end{equation}
where the supersymmetric conformal blocks $G_{\Delta,l}^\text{S}(z,\bar z)$ are given in appendix~\ref{blocks}, and implicitly depend on $n$ through $\Delta$. Note that equations~\eqref{susycrossing} and~\eqref{decomposition} imply that $G$ obeys the following crossing relation:
\begin{equation}
G(z,\bar{z})=-G(1-z,1-\bar{z}).
\label{Gcrossing}
\end{equation}

The free disconnected part of the four-point correlator can be computed in the abelian theory and corresponds to the following prepotential:
\begin{equation}
F(u,v)_{\text{free-disc}}=1+\frac{1}{u^{2}}+\frac{1}{v^{2}}.
\end{equation}
Decomposing this function according to~\eqref{decomposition} and computing the conformal block expansion of the piece encoding the non-protected operators according to~\eqref{eq:bigg} then gives the following formula for the leading contribution to the OPE coefficients: 
\begin{equation}
A_{n,l}^{(0)}=\frac{\left(l+2\right) \left(n+3\right)! \left(n+4\right)! \left(l+2 n+9\right) \left(l+2n+10\right) \left(l+n+5\right)! \left(l+n+6\right)!}{72 \left(2 n+5\right)! \left(2 l+2 n+9\right)!}.
\label{susyA0}
\end{equation}

\subsection{Recursion}

To derive recursion relations for the anomalous dimensions of the double-trace operators described above, we follow the same procedure as section~\ref{toy}. First expand the OPE data in $1/c$:
\begin{equation}
A_{n,l}=A_{n,l}^{(0)}+\frac{1}{c}A_{n,l}^{(1)}+...,\qquad\Delta=2n+l+8+\frac{1}{c}\gamma_{n,l}+...
\label{cexpansion}
\end{equation}
Focusing on the part of the prepotential which describes non-protected operators and expanding the crossing equation~\eqref{Gcrossing} to first order in $1/c$ then gives%
\footnote{Note that as in the toy model case, here $\gamma_{n,l}$ is an ``averaged anomalous dimension'', see footnote~\ref{f1}.}
\begin{equation}
\sum_{n,l\geq 0}\left[A_{n,l}^{(1)}\,G_{\Delta,l}^{\text{S}}(z,\bar{z})+\frac{1}{2}A_{n,l}^{(0)}\,\gamma_{n,l}\,\partial_{n}G_{\Delta,l}^{\text{S}}(z,\bar{z})\right]+\left(u\leftrightarrow v\right)=0.
\label{crossingexpansionsusy}
\end{equation}
In the supersymmetric case, the conformal blocks have the schematic form 
\begin{equation}\label{blockschematicsusy}
G^{\text{S}}_{\Delta,l}(z,\bar{z})\sim \sum u^{n}h_{\alpha}(z)h_{\beta}(\bar{z}),
\end{equation}
where
\begin{equation}\label{eq:hypergeomSUSY}
h_\beta(z)={}_2 F_1\left(\beta/2,\beta/2 -1,\beta,z\right)
\end{equation}
(see appendix~\ref{blocks} for details). Following the same reasoning described in the previous section, the term $\partial_{n}G_{\Delta,l}^{\text{S}}(z,\bar{z})$ in~\eqref{crossingexpansionsusy} gives a contribution proportional to $\log (z)$ and the analogous term in the cross channel will give $ \log(1-\bar{z})$, so we can isolate the terms containing anomalous dimensions by taking the limit $z \rightarrow 0$ and $\bar{z} \rightarrow 1$. In this case, the hypergeometrics depending on $\bar{z}$ and $1-z$ will give rise to $\log(1-\bar{z})$ and $\log(z)$ using the relation
\begin{equation}
h_\beta(\zb)=\log(1-\zb)\,(1-\zb)\,\tilde{h}_\beta(1-\zb)+\text{holomorphic at }\zb=1,
\label{hypertologsus}
\end{equation}
where
\begin{equation}
\tilde{h}_\beta(z)=\small{\frac{\Gamma(\beta)}{\Gamma(\beta/2) \Gamma(\beta/2-1)}} \normalsize {}_2F_1 \left(\beta/2+1,\beta/2,2,z\right).
\end{equation}

We thus consider the $\log(z)\log(1-\bar{z})$ coefficient of~\eqref{crossingexpansionsusy} in the limit $z \rightarrow 0$ and $\bar{z} \rightarrow 1$:
\begin{align}\label{eq:crossing}
&\sum_{n,l\geq 0}  A^{(0)}_{n, l}\,\gamma_{n, l}\left(\left.\partial_n G_{\Delta,\,l}^{\text S}
(z,\zb)\right)\right|_{\log z \log (1-\zb)}= \nonumber \\ 
&-\sum_{n,l\geq 0}   A^{(0)}_{n, l}\,\gamma_{n, l}\left(\left.\partial_n G_{\Delta,\,l}^{\text S}
(1-z,1-\zb)\right)\right|_{\log z \log (1-\zb)},
\end{align} 
into which we insert the (precise forms of) \eqref{blockschematicsusy} and \eqref{hypertologsus} to obtain sums of  terms involving $h_\alpha(z) \tilde h_\beta(1-\bar z)$ and $h_\alpha(1-\bar z) \tilde h_\beta(z)$. We then multiply this equation by
\begin{equation}
\frac{h_{-2q}(z)}{z^{q}\,(1-z)}\, \times \frac{h_{-2p}(1-\zb)}{(1-\zb)^{p}\,\zb},
\end{equation}
where $p$ and $q$ are arbitrary non-negative integers, and perform the contour integrals $\oint \frac{dz}{2\pi i} \oint \frac{d\zb}{2\pi i}$, which encircle $(z,\bar{z})=(0,1)$. Using the orthogonality relation proven in appendix~\ref{orthogonality} 
\begin{equation}
\delta_{m,m'}=\oint \frac{d z}{2\pi i}\,\frac{z^{m-m'-1}}{1-z}\,h_{2m+4}(z)\,h_{-2m'-2}(z),
\label{orthsusy}
\end{equation}
and defining
\begin{equation}
\mathcal{I}_{m,m'}=\oint \frac{d z}{2\pi i}\,\frac{(1-z)^{m-3}}{z^{m'-1}}\,\tilde{h}_{2m}(z)\,h_{-2m'}(z),
\end{equation}
finally leads to the following equation:
\begin{align}
0\,=&\sum_{l=0}^{L}\sum_{n=0}^{\infty} A^{(0)}_{n,l}\gamma_{n,l} \Bigl[P_{n,l}\left(\delta_{q,n}\mathcal{I}_{n+l+6,p+2}-\delta_{q,n+l+3}\mathcal{I}_{n+3,p+2}\right)\nonumber \\
& +Q_{n,l}\left(\delta_{q,n+2}\mathcal{I}_{n+l+6,p+2}-\delta_{q,n+l+3}\mathcal{I}_{n+5,p+2}\right)\nonumber \\ 
&+R_{n,l}\left(\delta_{q,n+l+2}\mathcal{I}_{n+4,p+2}-\delta_{q,n+1}\mathcal{I}_{n+l+5,p+2}\right)\nonumber\\
& +S_{n,l}\left(\delta_{q,n+l+4}\mathcal{I}_{n+4,p+2}-\delta_{q,n+1}\mathcal{I}_{n+l+7,p+2}\right)-\left(q\leftrightarrow p\right)\Bigr],
\label{susyrecursion}
\end{align}
where we have truncated the sum over spins and defined
\begingroup
\allowdisplaybreaks
\begin{align}
P_{n,l}=&\frac{l+1}{\left(n+3\right)\left(n+l+5\right)},\quad Q_{n,l}=\frac{\left(l+3\right)\left(n+5\right)\left(2n+l+8\right)}{4\left(2n+7\right)\left(2n+9\right)\left(n+l+5\right)\left(2n+l+10\right)},\nonumber \\
R_{n,l}=&\frac{l+3}{\left(n+3\right)\left(n+l+5\right)},\quad S_{n,l}=\frac{\left(l+1\right)\left(n+l+7\right)\left(2n+l+8\right)}{4\left(n+3\right)\left(2n+l+10\right)\left(2n+2l+11\right)\left(2n+2l+13\right)}.
\end{align}\\
\endgroup
As we explain in the next subsection and appendix~\ref{algorithm}, recursion relations for the anomalous dimensions are obtained from~\eqref{susyrecursion} by making appropriate choices for $p$ and $q$, and solutions are labelled by the spin truncation.  

\subsection{Solutions}\label{susysolutions}
In this subsection, we will describe solutions to the recursion relations for low-spin truncations and match them with results previously obtained in \cite{Heslop:2017sco}. For spin truncation $L=0$, setting $q=0$ in \eqref{susyrecursion} gives the following recursion relation in terms of $p$: 
\begin{equation}
\frac{1}{15} \mathcal{I}_{6,p+2} A^{(0)}_{0,0}\gamma_{0,0}  =\sum_{a=0}^{4}C_a A^{(0)}_{p-a,0}\gamma_{p-a,0},
\end{equation}
where
\begingroup
\allowdisplaybreaks
\begin{align}
C_0=&\,\frac{\mathcal{I}_{p+6,2}}{\left(p+3\right)\left(p+5\right)},\nonumber \\
C_1=&-\frac{3\,\mathcal{I}_{p+4,2}}{\left(p+2\right)\left(p+4\right)}-\frac{\left(p+3\right) \left(p+6\right)\mathcal{I}_{p+6,2}}{4 \left(p+2\right) \left(p+4\right) \left(2 p+9\right) \left(2 p+11\right)} ,\nonumber \\
C_2=&\,\frac{3\,\mathcal{I}_{p+2,2}}{\left(p+1\right)\left(p+3\right)} +\frac{3 \left(p+2\right)\mathcal{I}_{p+4,2}}{4\left(p+3\right)\left(2p+3\right)\left(2p+5\right)} ,\nonumber \\
C_3=&-\frac{\mathcal{I}_{p,2}}{p\left(p+2\right)}-\frac{3 \left(p+1\right)\mathcal{I}_{p+2,2}}{4\left(p+2\right)\left(2p+1\right)\left(2p+3\right)} ,\nonumber \\
C_4=&\,\frac{p\left(p+3\right)\mathcal{I}_{p,2}}{4 \left(p-1\right) \left(p+1\right) \left(2 p+3\right) \left(2 p+5\right)}.
\end{align}\\
\endgroup
This can be solved for all $\gamma_{n,0}$ in terms of $\gamma_{0,0}$ to give  
\begin{equation}
\gamma^{\text{spin}-0}_{n,0}=\gamma_{0,0}\,\frac{11 \left(n+1\right)_8 \left(n+2\right)_6}{2304000\left(2 n+7\right) \left(2 n+9\right) \left(2 n+11\right)},
\label{L=0}
\end{equation}
where $x_n=\Gamma (x+n)/\Gamma(x)$ and we divided by $A_{n,0}^{(0)}$, see~\eqref{susyA0}. It is interesting to compare this with the bosonic solution \eqref{bos0} for $\Delta_0=4$:
\begin{equation}
\left(\gamma_\text{bos}\right)_{n,0}^{\text{spin-0}}=\left(\gamma_\text{bos} \right)_{0,0}\,\frac{35 \left(n+1\right)_4 \left(n+2\right)_2}{96 \left(2 n+3\right) \left(2 n+5\right) \left(2 n+7\right)}.
\end{equation}
Similarly, following the procedure described in section~\ref{toy} and appendix~\ref{algorithm} we obtain solutions for $L=2$ in terms of three unfixed parameters $\left\{\gamma_{0,0},\,\gamma_{0,2},\,\gamma_{1,2}\right\}$, which are given in appendix~\ref{SolsSpin2}.

More generally, for spin truncation $L$, the solution will depend on $(L+2)(L+4)/8$ free parameters, in agreement with the counting of solutions in section~\ref{toy}. Moreover, our results for the anomalous dimensions agree with those obtained in \cite{Heslop:2017sco}, which deduced solutions to the crossing equations whose conformal block expansions are truncated in spin. In particular, the anomalous dimensions in~\eqref{L=0} can be obtained from the conformal block expansion of
\begin{equation}
F^{\text{spin-0}}(u,v)=C^{(0)} \lambda^2 u v \bar{D}_{5755}(u,v),
\end{equation}
where the coefficient $C^{(0)}$ is unfixed. Decomposing $F^{\text{spin-0}}$ according to~\eqref{decomposition} and performing the conformal block expansion according to~\eqref{crossingexpansionsusy} gives the anomalous dimensions in~\eqref{L=0} if we choose the free parameter to be
\begin{equation}
\gamma_{0,0}=-\frac{7200\,C^{(0)}}{77}.
\end{equation}
Following the holographic arguments of \cite{Heemskerk:2009pn,Alday:2014tsa,Heslop:2017sco}, which were reviewed in section~\ref{toy}, $F^{\text{spin-0}}$ should arise from an $\mathcal{R}^4$ correction to supergravity in $AdS_7 \times S^4$, where $\mathcal{R}$ is the Riemann tensor. This can be seen by noting that in the large-$n$ limit the anomalous dimensions scale like $n^6$ times the anomalous dimensions obtained in the supergravity approximation, indicating that the corresponding interaction vertex has six more derivatives than supergravity. 

Note that $F$ is a prepotential from which many four-point component correlators (corresponding to different choices of $Y_i$) are obtained by applying a differential operator according to \eqref{prepot}. This differential operator can be rewritten in terms of $u,v$ derivatives and so if the prepotential is expressed in terms of $\bar{D}$ functions then so will all the component correlators. Whilst this does not prove that the prepotential can always be expressed in terms of $\bar{D}$ functions, this property holds in all the examples we have considered, and it is natural to  conjecture that it should hold in general. A similar conjecture was made in \cite{Rastelli:2017ymc} for four-point correlators of more general half-BPS operators in the supergravity approximation. 

For the $L=2$ spin truncation, \cite{Heslop:2017sco} found the following solutions to the crossing equation:
\begingroup
\allowdisplaybreaks
\begin{align}
&F_4^\text{spin-2}(u,v)=2\, C_4^{(2)}\,\lambda^2 u v \left(\bar{D}_{6776}(u,v)+\bar{D}_{7676}(u,v)+\bar{D}_{7766}(u,v)\right),\\
&F_6^{\text{spin-2}}(u,v)=6\, C_6^{(2)}\, \lambda^2 u v \bar{D}_{7777}(u,v),
\end{align}
\endgroup
where the coefficients $C_{4,6}^{(2)}$ are unfixed and the subscripts indicate the number of additional derivatives compared to the bulk interaction vertex associated with the $L=0$ solution. The first solution corresponds to a $D^4 \mathcal{R}^4$ correction and the second one corresponds to a $D^6 \mathcal{R}^4$ correction to supergravity in $AdS_7 \times S^4$, which can be read off from the large-twist behaviour of the corresponding anomalous dimensions, as described in appendix~\ref{SolsSpin2}. The anomalous dimensions of these two solutions are reproduced from the general solution in appendix~\ref{SolsSpin2} for the choices
\begingroup
\allowdisplaybreaks
\begin{align}
\label{15der}
\left\{ \gamma_{0,0},\gamma_{0,2},\gamma_{1,2}\right\} _{4}&=C_4^{(2)}\left\{ -\frac{5\times72000}{1001},\frac{80640}{1859},\frac{5\times150528}{2431}\right\}, \\
\left\{ \gamma_{0,0},\gamma_{0,2},\gamma_{1,2}\right\} _{6}&=C_6^{(2)}\left\{ \frac{54\times72000}{1001},-\frac{3\times80640}{1859},-\frac{33\times150528}{2431}\right\}. 
\end{align}\\
\endgroup
Note that in both cases, $\gamma_{0,0}$ has the opposite sign of $\gamma_{0,2}$ and $\gamma_{1,2}$, in contrast to what we found for the toy model in~\eqref{bosparams1} and~\eqref{bosparams2}, where all three parameters had the same sign. Furthermore, the ratio of the coefficients $\gamma_{0,2}$ and $\gamma_{1,2}$ in \eqref{15der} can be fixed from large-twist behaviour as we explain in appendix~\ref{SolsSpin2}. The attached Mathematica notebook {\tt{6drecursion.nb}} can be used to solve the recursion relations up to any desired twist and spin truncation.

Although solutions to the recursion relations have unfixed coefficients, it is possible to deduce their leading $1/c$-dependence using holographic reasoning, as described in \cite{Heslop:2017sco}. First note that since we solve the recursion relations by truncating in spin, this restricts to contact interactions in the bulk (interactions involving bulk-to-bulk propagators will not truncate in spin). The effective action then has the schematic form 
\begin{equation}
\mathcal{L}\sim\frac{1}{G_{N}}\left[\left(\partial\phi\right)^{2}+\sum_{D} l_P^{D-2} \partial^{D}\phi^{4}\right],
\end{equation}
where $\phi$ represents a graviton field, $G_N$ is Newton's constant, and the Planck length $l_P$ is inserted by dimensional analysis. After rescaling the graviton by $\sqrt{G_N}$ in order to have canonical kinetic terms, the four-point interactions will acquire a factor of $G_N \sim 1/c$  (this is the origin of the $1/c$ in \eqref{cexpansion}). Recalling that $G_N \sim l_P^{9}$ in eleven dimensions, we see that a four-point contact interaction with $D$ derivatives must therefore have coefficient $G_N l_P^{D-2}\sim c^{-(D+7)/9}$. Moreover, the number of derivatives in a contact interaction can be read off from the large-twist behaviour of the corresponding solution to the crossing equations~\cite{Heemskerk:2009pn}. In particular, if the anomalous dimensions of the solution scale like $n^{\alpha}$, then the corresponding bulk interaction must have $D=(\alpha-5)+2=\alpha-3$ derivatives (recalling that anomalous dimensions scale like $n^5$ in the supergravity approximation). 

In summary, a solution whose anomalous dimensions scale like $n^\alpha$ must have a coefficient $c^{-(\alpha+4)/9}$. For example, the spin-0 solution in \eqref{L=0} will have a coefficient of $c^{-5/3}$ and spin-2 solutions which scale like $n^{15}$ and $n^{17}$ will have coefficients of $c^{-19/9}$ and $c^{-7/3}$, respectively. Note that similar reasoning applies to conformal field theories with string theory duals, like $\mathcal{N}=4$ SYM with any fixed finite value of the string coupling. In that case, a contact interaction with $D$ derivatives will have a coefficient of $G_N \alpha'^{(D-2)/2}$, where $\alpha'$ is related to the square of the string length. Writing this prefactor in terms of the central charge and string coupling, and fixing the latter at some finite value  
will then give an expansion in $1/c$ analogous to M-theory.    

\section{Conclusion} \label{conclusion}

In this paper, we derive recursion relations for anomalous dimensions of double-trace operators in the 6d $(2,0)$ theory. Given that no Lagrangian description is presently known for this model, our strategy is to use superconformal and crossing symmetry of four-point correlators of stress tensor multiplets. In particular, we expand the crossing equation to first order in the inverse central charge and then take a certain limit of the conformal cross ratios to isolate the terms containing anomalous dimensions. Recursion relations then follow from truncating the conformal block expansion in spin and taking inner products of the resulting equation with certain hypergeometric functions. These recursion relations can then be solved to obtain anomalous dimensions for arbitrary twist and spin, reproducing the results for low spin truncations previously obtained in \cite{Heslop:2017sco}. As a warm-up, we derive analogous recursion relations in a toy model corresponding to an abstract bosonic 6d CFT, and match the results with the conformal block expansion of Witten diagrams in $AdS_7$, confirming the holographic arguments of \cite{Heemskerk:2009pn}. Moreover, these recursion relations are easily implemented on a computer, and we attach the Mathematica file {\tt{6drecursion.nb}} for computing anomalous dimensions in both the bosonic and supersymmetric theories to any desired twist and spin truncation. We note that this method for extracting anomalous dimensions is much more efficient than extracting them using a conformal block expansion of a known four-point function.

The anomalous dimensions are physically significant because they encode higher- derivative corrections to supergravity in $AdS_7 \times S^4$. In particular, they appear in the conformal block expansion of solutions to the crossing equations which reduce to scattering amplitudes of the low energy effective action for M-theory in the flat space limit. The number of derivatives in each term of the effective action can be read off from the large-twist behaviour of the corresponding anomalous dimensions. Moreover, the coefficients of these higher-derivative terms correspond to free parameters of the solutions to the recursion relations and are therefore not determined by this approach. In the flat space limit, the coefficients of the $\mathcal{R}^4$ and $D^6 \mathcal{R}^4$ terms in the M-theory effective action have been deduced by uplifting string theory amplitudes (note that the $D^4 \mathcal{R}^4$ term vanishes in 11 dimensions) \cite{Green:1997di,Green:2005ba}, but the coefficient of the $D^8 \mathcal{R}^4$ term (which arises from a truncated spin-4 solution in our approach) is unknown. It would therefore be desirable to develop methods for fixing these coefficients using CFT techniques. 

A strategy for doing so was proposed in \cite{Chester:2018dga}, and used to fix the coefficient of the $\mathcal{R}^4$ term and argue that the $D^4 \mathcal{R}^4$ term vanishes. This was achieved by applying the chiral algebra conjecture in \cite{Beem:2014kka} to four-point correlators of the form $\left\langle kkkk\right\rangle $ with $k=3$, where $k$ refers to a half-BPS scalar operator in the $k$-index symmetric traceless representation of the $SO(5)$ R-symmetry group with scaling dimension $2k$ (note that $k=2$ is the case considered in this paper). It would therefore be interesting to find truncated spin solutions to the crossing equations for higher-charge correlators, formulate recursion relations for the anomalous dimensions in their conformal block expansions, and ultimately fix the coefficients of higher-derivative terms in the M-theory effective action \footnote{Correlators of the form $\left\langle kkkk\right\rangle $ and $\left\langle n+k,n-k,k+2,k+2\right\rangle $ were computed in the supergravity approximation in \cite{Rastelli:2017ymc,Zhou:2017zaw}. Moreover new solutions to the conformal Ward identities in Mellin space have been found for $\left\langle kkkk\right\rangle$ with $k=2,3$ in \cite{Chester:2018dga}, so it would be interesting to see how those methods are related to the ones developed in this paper.}. It would also be interesting to consider mixed correlators such as $\left\langle ppqq \right\rangle $, although our method does not immediately extend to such correlators because they do not have the required crossing symmetry. Since the conformal blocks for higher-charge correlators appear to be much simpler in 4d \cite{Dolan:2003hv,Doobary:2015gia}, it may be instructive to first carry out the analysis described above for 4d $\mathcal{N}=4$ SYM (for which a chiral algebra description was also proposed in \cite{Beem:2013sza}), and use it to deduce terms in the effective action for IIB string theory in $AdS_5 \times S^5$.

It would also be very interesting to explore the loop expansion in M-theory on $AdS_7\times S^4$ using conformal bootstrap techniques, following on from the recent success on $AdS_5\times S^5$~\cite{Aprile:2017bgs,Aprile:2017qoy,Aprile:2017xsp,Aprile:2018efk,Alday:2017vkk,Alday:2017xua,Alday:2018kkw,Alday:2018pdi,Caron-Huot:2018kta}.

\begin{center}
\textbf{Acknowledgements}
\end{center}
TA is supported by a Durham Doctoral Studentship, AL by the Royal Society as a Royal Society University Research Fellowship holder, and PH by an STFC Consolidated Grant ST/P000371/1.

\appendix

\section{Conformal Blocks} \label{blocks}

Conformal blocks for four-point correlators of scalar operators of arbitrary scaling dimensions $\Delta_i$, $i=1,...,4$, in any even dimension were derived by Dolan and Osborn in \cite{Dolan:2003hv}. In 6d, the blocks are given by
\begin{align}
G^{\text{DO}}&\left(\Delta,l,\Delta_{12},\Delta_{34}\right)=\nonumber\\
&\mathcal{F}_{00}-\frac{l+3}{l+1}\mathcal{F}_{-11}\nonumber\\
&-\frac{\Delta-4}{\Delta-2}\,\frac{\left(\Delta+l-\Delta_{12}\right)\left(\Delta+l+\Delta_{12}\right)\left(\Delta+l+\Delta_{34}\right)\left(\Delta+l-\Delta_{34}\right)}{16\left(\Delta+l-1\right)\left(\Delta+l\right)^2\left(\Delta+l+1\right)}\mathcal{F}_{11}\nonumber \\
&+\frac{\left(\Delta-4\right)\left(l+3\right)}{\left(\Delta-2\right)\left(l+1\right)}\nonumber\\
&\times \frac{\left(\Delta-l-\Delta_{12}-4\right)\left(\Delta-l+\Delta_{12}-4\right)\left(\Delta-l+\Delta_{34}-4\right)\left(\Delta-l-\Delta_{34}-4\right)}{16\left(\Delta-l-5\right)\left(\Delta-l-4\right)^2\left(\Delta-l-3\right)}\mathcal{F}_{02}\nonumber \\
&+2\left(\Delta-4\right)\left(l+3\right)\frac{\Delta_{12}\Delta_{34}}{\left(\Delta+l\right)\left(\Delta+l-2\right)\left(\Delta-l-4\right)\left(\Delta-l-6\right)}\mathcal{F}_{01},
\end{align}
where $(\Delta,l)$ are the scaling dimension and spin of a primary operator in the conformal block expansion, $\Delta_{ij}=\Delta_{i}-\Delta_{j}$, and  
\begin{align}
\mathcal{F}_{ab}=&\frac{(z\bar{z})^{\tfrac{1}{2}(\Delta-l)}}{\lambda^{3}}\Big\{z^{l+a+3}\zb^{b}\nonumber \\
&\times {}_{2}F_{1}\left(\tfrac{1}{2}(\Delta+l-\Delta_{12})+a,\tfrac{1}{2}(\Delta+l+\Delta_{34})+a;\Delta+l+2a,z\right)\nonumber \\
&\times {}_{2}F_{1}\left(\tfrac{1}{2}(\Delta-l-\Delta_{12})-3+b,\tfrac{1}{2}(\Delta-l+\Delta_{34})-3+b;\Delta-l-6+2b;\bar{z}\right)\nonumber \\
& -z\leftrightarrow \bar{z} \Big\}.
\end{align}
For the toy model analysed in section~\ref{toy}, the blocks are given by 
\begin{equation}
G_{\Delta,\,\l}^{\text{B}}(z,\bar{z})=\left(l+1\right)G^{\text{DO}}(\Delta,l,0,0),
\end{equation}
where $\Delta=2n+l+2\Delta_0+\mathcal{O}(1/c)$. Moreover, for the 6d $(2,0)$ theory analysed in section~\ref{susy}, the blocks are given by \cite{Heslop:2004du,Beem:2015aoa}
\begin{equation}
G_{\Delta,\,\l}^{\text{S}}(z,\bar{z})=\frac{4\left(l+1\right)}{\left(l+2\right)^{2}-\Delta^{2}}\,\frac{\lambda^{3}}{u^{5}}\,G^{\text{DO}}(\Delta+4,l,0,-2),
\end{equation}
where $\Delta=2n+l+8+\mathcal{O}(1/c)$ with $n \geq 0$. 

\section{Orthogonality of Hypergeometrics}\label{orthogonality}
In this appendix we derive orthogonality relations for hypergeometric functions used in this paper, explicating a brief argument in~\cite{Heemskerk:2009pn} which then allows us to obtain a new case relevant for the supersymmetric 6d theory. Our starting point will be the differential operator%
\footnote{ Note that this operator is closely related to the conformal Casimir. In $d$ dimensions this is \cite{Dolan:2003hv}
	\begin{align}
	D_\epsilon\,=\,&z^2\,(1-z)\,\partial_z^2+\zb^2\,(1-\zb)\,\partial_{\zb}^2-(a+b+1)\,(z^2\partial_z+\zb^2\partial_{\zb})\nonumber \\
	&-a\,b\,(z+\zb)+\epsilon\,\frac{z\,\zb}{z-\zb}\left((1-z)\,\partial_z-(1-\zb)\,\partial_{\zb}\right),
	\end{align}
	where $a,b$ are arbitrary constants and $\epsilon=d-2$. The non-interacting part (i.e $\epsilon$-independent part) reduces to $D_z+D_{\bar{z}}$.}
\begin{equation}
D_z=z^2\,(1-z)\,\partial_z^2-(a+b+1)z^2\partial_z-a\,b\,z.
\label{dz}
\end{equation}
This operator has eigenfunctions satisfying
\begin{equation}\label{eq:EVequ}
D_{z}H_{m}(z)=m(m-1)H_{m}(z),
\end{equation}
where
\begin{equation}
H_{m}(z)=z^{m}{}_{2}F_{1}(m+a,m+b;2m;z).
\label{hypergeometricb}
\end{equation}
First consider $a=b=0$. In this case, the differential operator in~\eqref{dz} reduces to $D_z=z^2\partial_z\,(1-z)\partial_z$. Let us look at the object $H_{m}H_{1-m'}$ (we will omit the arguments $(z)$ in the following). Using the symmetry of the differential operator $D_z$, after integrating by parts twice and using~\eqref{eq:EVequ} we find that
\begin{align}
0=&\oint \frac{dz}{2\pi i}\frac{1}{z^2}\left[\left(D_{z}H_{m}\right)H_{1-m'}-H_{m}\left(D_{z}H_{1-m'}\right)\right]\nonumber \\
=&[m(m-1)-m'(m'-1) ]\oint \frac{dz}{2\pi i}\frac{1}{z^2}H_{m}H_{1-m'},
\end{align}
where the contour encircles the origin. It follows that $H_m$ and $H_{1-m'}$ are orthogonal with respect to the inner product defined above if $m \neq m'$. Plugging in~\eqref{hypergeometricb} and shifting $(m,m')$ to $(m+2,m'+2)$ then implies the inner product in~\eqref{boseorthogonal}, where we fix the normalisation by noting that $_{2}F_{1}(\alpha,\beta,\gamma,z)=1+\mathcal{O}\left(z\right)$ and evaluating the residue at $z=0$. This was first obtained in~\cite{Heemskerk:2009pn}.

Next, consider $a=0,\ b=-1$, in which case~\eqref{dz} reduces to $D_z=z^2\,(1-z)\,\partial_z^2$. Following the same arguments as above we find that
\begin{align}
0=&\oint\frac{dz}{2\pi i}\frac{1}{z^{2}(1-z)}\left[\left(D_{z}H_{m}\right)H_{1-m'}-H_{m}\left(D_{z}H_{1-m'}\right)\right]\nonumber \\
=&[m(m-1)-m'(m'-1)]\oint\frac{dz}{2\pi i}\frac{1}{z^{2}(1-z)}H_{m}H_{1-m'},
\end{align}
so $H_m$ and $H_{1-m'}$ are orthogonal with respect to the inner product defined above if $m \neq m'$. Plugging in~\eqref{hypergeometricb} and shifting $(m,m')$ to $(m+2,m'+2)$ then proves the inner product in~\eqref{orthsusy}, where the normalisation is once again fixed by evaluating the residue at $z=0$. 

\section{Solving the Recursion Relations} \label{algorithm}
Recursion relations for the anomalous dimensions of double-trace operators in the 6d toy model and $(2,0)$ theory are encoded in \eqref{eq:BosRecursion} and \eqref{susyrecursion}, respectively, and are obtained by specifying a spin truncation $L$ and making appropriate choices of non-negative integers $p$ and $q$. The general algorithm for solving the recursion relation for a general spin-$L$ truncation is as follows: 
\begin{itemize}
\item For each $1 \leq p \leq L/2$, write down the equations for $0 \leq q \leq p-1$.
\item Solve these equations for $\gamma_{p,l}$ with $0 \leq l \leq 2p -2$ in terms of $\gamma_{p',l'}$ with $(p' \leq p-1,\,l' \leq L)$ and $(p'=p,\, 2p \leq l' \leq L)$.
\item For each $p \geq L/2+1$, write down the equations for $0 \leq q \leq L/2$.
\item Solve these equations for $\gamma_{p,l}$ with $0\leq l \leq L$ in terms of $\gamma_{p',l'}$ with $(p' \leq p-1,\,l'\leq L)$.
\end{itemize}
In the end, this algorithm will give all $\gamma_{n,l}$ with $l \leq {\text{min}} (2n-2,L)$ in terms of all $\gamma_{n',l'}$ with $2n' \leq l'\leq L$, which correspond to $(L+2)(L+4)/8$ free parameters as depicted in Figure 3 of \cite{Heemskerk:2009pn}. This algorithm is implemented in the attached Mathematica notebook {\tt{6drecursion.nb}} by generating all the free parameters for a given $L$, writing down the equations for every $p \geq 1$ and $0 \leq q \leq \min(p-1,L/2)$, replacing $(L+2)(L+4)/8$ of the anomalous dimensions by the free parameters, and solving these equations for the remaining anomalous dimensions.

\section{L=2 Solutions}\label{SolsSpin2}
In this appendix we give the solutions to the recursion relations for anomalous dimensions with spin truncation $L=2$ in the toy model and $(2,0)$ theory. For the toy model, we find the following solutions for general scaling dimension $\Delta_0$:
\begin{align}
\gamma^{ \text{spin-2}}_{n,0}\left(\Delta_0\right)=&\frac{\gamma^{\text{spin-0}}_{n,0}\left(\Delta_0\right)}{\gamma_{0,0}}\left(\gamma_{0,0}+\gamma_{0,2}\, f_1\left(n,\Delta_0\right)+\gamma_{1,2} \,f_2\left(n,\Delta_0\right)\right),\label{Spin2Bos1} \\
\gamma^{\text{spin-2}}_{n,2}\left(\Delta_0\right)=&-\frac{\gamma^{\text{spin-0}}_{n,0}\left(\Delta_0\right)}{\gamma_{0,0}}\frac{\left(2 \Delta_0 +1\right)^2 \left(2 \Delta_0 +3\right) \left(n-1\right) \left(n+3\right) \left(n+4\right) }{4\,\left(\Delta_0 -1\right) \Delta_0 ^4 \left(\Delta_0 +1\right)^2 }\nonumber \\
&\times   \frac{\left(\Delta_0 +n\right) \left(\Delta_0 +n+1\right) \left(2 \Delta_0 +n\right)\left(2 \Delta_0 +n-3\right) \left(2 \Delta_0 +n-2\right)}{\left(2\Delta_0-3\right)\left(2 \Delta_0 +2 n+1\right) \left(2 \Delta_0 +2 n+3\right)} \nonumber \\
&\times \left(\gamma_{0,2}-\gamma_{1,2}\,\frac{4 \Delta_0  \left(2 \Delta_0 +3\right) \left(2 \Delta_0 +5\right) n \left(2 \Delta_0 +n-1\right)}{\left(\Delta_0 +1\right) \left(\Delta_0 +2\right)^2 \left(2 \Delta_0 -1\right) \left(2 \Delta_0 +1\right) \left(n-1\right) \left(2 \Delta_0 +n\right)}\right),\label{Spin2Bos2}
\end{align}
where
\begin{align}
f_1\left(n,\Delta_0\right)=&\,\frac{\left(2 \Delta_0 +1\right)^2 \left(2 \Delta_0 +3\right) n \left(2 \Delta_0 +n-3\right)}{\left(\Delta_0 -1\right) \Delta_0 ^4 \left(\Delta_0 +1\right)^2 \left(2 \Delta_0 -3\right) \left(2 \Delta_0 +2 n-7\right) \left(2 \Delta_0 +2 n+1\right)}\nonumber \\
&\times \Biggl( 5 n^6+15 \left(2 \Delta_0 -3\right) n^5 +\left(89 \Delta_0 ^2-161 \Delta_0 +127\right) n^4 +\left(2 \Delta_0 -3\right)\nonumber \\
&\times \left(78 \Delta_0 ^2-22 \Delta_0 +29\right) n^3+2 \left(82 \Delta_0 ^4-143 \Delta_0 ^3-107 \Delta_0 ^2+117 \Delta_0 -39\right) n^2\nonumber \\
&+\left(2 \Delta_0 -3\right) \left(48 \Delta_0 ^4-14 \Delta_0 ^3-215 \Delta_0 ^2-33 \Delta_0 -6\right) n\nonumber \\
&+\frac{6\left(\Delta_0 -1\right) \Delta_0 ^2 \left(2 \Delta_0 -7\right) \left(4 \Delta_0 ^3+12 \Delta_0 ^2+5 \Delta_0 -1\right)}{2 \Delta_0 +1}\Biggr),\nonumber
\end{align}
\begin{align}
f_2\left(n,\Delta_0\right)=&\frac{\left(2 \Delta_0 +1\right) \left(2 \Delta_0 +3\right)^2 \left(2 \Delta_0 +5\right) n \left(2 \Delta_0 +n-3\right)}{\left(3-2 \Delta_0 \right) \left(\Delta_0 -1\right) \Delta_0 ^3 \left(\Delta_0 +1\right)^3 \left(\Delta_0 +2\right)^2 \left(2 \Delta_0 -1\right) \left(2 \Delta_0 +2 n-7\right) }\nonumber \\
&\times\frac{1}{\left(2 \Delta_0 +2 n+1\right)} \Bigl( 20 n^6+60 \left(2 \Delta_0 -3\right) n^5+\left(4 \Delta_0  \left(89 \Delta_0 -199\right)+508\right) n^4\nonumber \\
&+4 \left(2 \Delta_0 -3\right) \left(78 \Delta_0 ^2-98 \Delta_0 +29\right) n^3+8 \left(\Delta_0  \left(2 \Delta_0  \left(\Delta_0  \left(41 \Delta_0 -131\right)+104\right)\right.\right.\nonumber \\
&\left.\left.-27\right)-39\right) n^2+4 \left(2 \Delta_0 -3\right) \left(\Delta_0  \left(\Delta_0  \left(4 \Delta_0  \left(12 \Delta_0 -25\right)-41\right)+21\right)-6\right) n\nonumber \\
&+ 24 \left(\Delta_0 -1\right) \Delta_0 ^2 \left(\Delta_0 +1\right) \left(2 \Delta_0 -7\right) \left(2 \Delta_0 -1\right)\Bigr).
\end{align}
As explained in subsection~\ref{toysolutions}, the large-$n$ limit of the anomalous dimensions corresponds to the number of derivatives in the bulk interactions. Since~\eqref{Spin2Bos1} and~\eqref{Spin2Bos2} scale like $n^{9}$ for large $n$, while the spin-0 solution in \eqref{bos0} scales like $n^{3}$, we see that these solutions correspond to six-derivative interactions. We also expect to have anomalous dimensions corresponding to four-derivative interactions, which scale like $n^{7}$ in the large-$n$ limit. We indeed find such solutions for the following choice of free parameters:
\begin{equation}
\gamma_{1,2}=\frac{\left(\Delta_0 +1\right) \left(\Delta_0 +2\right)^2 \left(2 \Delta_0 -1\right) \left(2 \Delta_0 +1\right) }{4 \Delta_0  \left(2 \Delta_0 +3\right) \left(2 \Delta_0 +5\right)}\,\gamma_{0,2},
\end{equation}
which is deduced by imposing that the large-$n$ limit of the last line in~\eqref{Spin2Bos2} vanishes. Note that the solution in \eqref{bosparams1} is consistent with this constraint. More generally, for a spin-$L$ solution one can deduce $L/2$ constraints on the coefficients (corresponding to bulk interactions with the number of derivatives ranging from $2L$ to $3L-2$ in intervals of 2) by analysing the large-twist limit. Unconstrained coefficients then encode the freedom to add solutions with lower spin or subleading large-twist behaviour.

In the 6d $(2,0)$ theory, we find the following solutions for spin truncation $L=2$:
\begin{align}
\gamma^{\text{spin-2}}_{n,0}=&\frac{\gamma^{\text{spin-0}}_{n,0}}{\gamma_{0,0}}\left(\gamma_{0,0}+\gamma_{0,2}\,f_1\left(n\right)+\gamma_{1,2}\,f_2\left(n\right)\right),\label{Spin2Susy1} \\
\gamma^{\text{spin-2}}_{n,2}=&-\frac{\gamma_{n,0}^{\text{spin-0}}}{\gamma_{0,0}}\,\frac{845 \left(n-1\right) \left(n+5\right) \left(n+6\right) \left(n+8\right) \left(n+9\right)^2 \left(n+10\right) \left(n+12\right) }{4064256 \left(2 n+13\right) \left(2 n+15\right)}\nonumber \\ &\times\left(\gamma_{0,2}-\gamma_{1,2}\,\frac{51 n \left(n+11\right)}{364 \left(n-1\right) \left(n+12\right)}\right),\label{Spin2Susy2} 
\end{align}where
\begingroup
\allowdisplaybreaks
\begin{align}
f_1\left(n\right)=&\,\frac{325n \left(n+9\right) \left(13 n^6+351 n^5+6201 n^4+64233 n^3+385476 n^2+1251666 n+1512620\right)}{1016064 \left(2 n+5\right) \left(2 n+13\right)},\nonumber\\
f_2\left(n\right)=&-\frac{1105 n \left(n+9\right) \left(5 n^6+135 n^5+2157 n^4+20601 n^3+117468 n^2+370494 n+441700\right)}{9483264 \left(2 n+5\right) \left(2 n+13\right)}.
\end{align}\\
\endgroup
In the large-$n$ limit,~\eqref{Spin2Susy1} and~\eqref{Spin2Susy2} scale like $n^{17}$. Since the spin-0 solution in \eqref{L=0} scales like $n^{11}$, it follows that the corresponding bulk interactions have six more derivatives. We also expect to find solutions corresponding to interactions with four more derivatives, which scale like $n^{15}$ for large $n$. We obtain these solutions for the choice of free parameters
\begin{equation}
\gamma_{1,2}=\frac{364}{51}\,\gamma_{0,2},
\end{equation}
which comes from imposing that the last line of~\eqref{Spin2Susy2} vanishes in the large-$n$ limit. The solution in \eqref{15der} is consistent with this constraint.

\end{document}